\documentstyle[twoside,fleqn,espcrc2,psfig]{article}

\newcommand{\AmS}{{\protect\the\textfont2
  A\kern-.1667em\lower.5ex\hbox{M}\kern-.125emS}}
\newcommand{\mysection}[1]{\vspace*{-2mm}\section{#1}\vspace*{-2mm}}
\title{Scalar and axial vector matrix elements of proton in quenched
QCD: Calculation of both connected and
      disconnected contributions
       \thanks{presented by Y.~Kuramashi}}
\author{ M.~Fukugita
         \address{Yukawa Institute for Theoretical Physics,
         Kyoto University, Kyoto 606, Japan }%
         , Y.~Kuramashi
         \address{National Laboratory for High Energy Physics(KEK),
         Tsukuba, Ibaraki 305, Japan}%
         , M.~Okawa$^{\:\:{\rm b}}$
         and
         A.~Ukawa
         \address{Institute of Physics,  University of Tsukuba,
         Tsukuba, Ibaraki 305, Japan }}

\begin{document}

\begin{abstract}

$\pi$-$N$ $\sigma$ term and proton axial vector matrix elements are calculated
including disconnected contributions with a variant wall source method using
the Wilson quark action at $\beta=5.7$ in quenched QCD. For the $\sigma$ term,
we find $\sigma_{disc}/\sigma_{conn}=2.35(46)$ and
$\sigma=44(6)$MeV--60(9)MeV.  The fraction of proton spin
withXcarried by quarks is
$\Delta\Sigma=\Delta{u}+\Delta{d}+\Delta{s}=+0.638(54)-0.347(46)-0.109(30)=
+0.18(10)$. \end{abstract}

\maketitle

\section{Introduction}
\vspace*{-2mm}
Interest in the pion-nucleon $\sigma$ term and the flavor singlet axial vector
matrix element of proton stems from the experimental
indication\cite{koch,protonspin} that sea quark contributions appear to be
significant in these matrix elements. Lattice QCD studies of these
quantities\cite{previous}, however, have been stagnant because of the
difficulty of calculating disconnected
contributions\cite{z2noise,fullw,fullks,axialff}.  Here we report on our
results obtained with the method of wall source without gauge fixing,
previously applied successfully to evaluate hadron scattering
lengths\cite{scattering} and the $\eta^\prime$ meson mass\cite{eta}, both
sharing a similar computational problem.

\begin{table}[t]
\setlength{\tabcolsep}{0.2pc}
\caption{Number of configurations ($N_c)$ and hadron mass results on
an $L^3\times 20$ lattice at $\beta=5.7$.} \label{table:one}
\begin{tabular}{llllll}
\hline
$K$  & $L$ & $N_c$ & $m_\pi a$ & $m_\rho a$ & $m_Na$ \\
\hline
0.1600 & 12 & 300 & 0.6876(31) & 0.8053(39) & 1.2957(79)\\
       & 16 &	260 & 0.6873(24) &	0.8021(29) &	1.2900(60)\\
0.1640 & 12 & 300 & 0.5080(37) & 0.6865(53) & 1.080(10)\\
       & 16	& 260	& 0.5080(29)	& 0.6822(38) &	1.0738(80)\\
0.1665 & 12 & 400 & 0.3663(44) & 0.6086(75) & 0.926(13)\\
       & 16 & 260 &	0.3674(39)	& 0.6085(58) &	0.915(11)\\
\hline
\end{tabular}
\vspace*{-5mm}
\end{table}

Our calculations are carried out in quenched QCD at $\beta=5.7$
Wilson quark action on an $L^3\times 20$ lattice. Number of
configurations, heat-bath generated at 1000 sweep intervals, are tabulated in
Table~\ref{table:one} for each spatial size $L$  and the hopping parameter
$K$, together with results for hadron masses. The nucleon matrix elements are
extracted in the standard way from the nucleon three-point function
divided by the nucleon propagator.
The connected contribution is obtained with the source method.  For the
disconnected contribution we use quark propagators with unit source for all
space-time lattice points evaluated without gauge fixing\cite{eta}.
Tadpole-improved one-loop expressions are employed for the $Z$ factor with
$\alpha_{\overline{MS}}(1/a)=0.2207$.  Errors  are estimated by the single
elimination jackknife procedure.

\begin{figure}[t]
\psfig{file=fullsigma.epsf,height=6cm,width=7.5cm,angle=-90}
\vskip -1.0cm
\caption{Comparison of results for the full $\sigma$ term.}
\label{fig:sigmasummary}
\vspace*{-9mm}
\end{figure}

\mysection{Scalar matrix elements and $\pi$-$N$ $\sigma$ term}

\begin{table*}[htb]
\setlength{\tabcolsep}{0.2pc}
\newlength{\digitwidth} \settowidth{\digitwidth}{\rm 0}
\catcode`?=\active \def?{\kern\digitwidth}
\caption{Nucleon scalar matrix elements as a function of $K$ for up and down
quarks.}
\label{table:two}
\begin{tabular*}{\textwidth}{@{}l@{\extracolsep{\fill}}lllllll} \hline $K$  &
$L$& \multicolumn{2}{c}{$<N|\bar uu+\bar dd|N>$} & $<N|\bar ss|N>$ & $F_S$ &
$D_S$ \\
     &    & conn. & disc.  \\
\hline
0.1600 & 12 & 2.323(15) & 3.56(76)  & 2.09(25) & 0.749(5)   & $-$0.074(2)\\
       & 16 &	2.326(17) & 3.02(98)  & 1.78(32) &	0.749(7)
& $-$0.075(6)\\
0.1640 & 12 & 2.413(30) & 4.58(92)  & 2.36(29) & 0.762(10)  & $-$0.126(6)\\
       & 16	& 2.378(30)	& 4.1(1.2)	 & 2.10(40) & 0.748(12)
& $-$0.128(11)\\
0.1665 & 12 & 2.693(81) & 5.1(1.1)  & 2.66(35) & 0.831(25)  & $-$0.200(16)\\
       & 16 & 2.565(73) &	5.6(1.6)	 & 2.70(51) & 0.783(26)
& $-$0.204(21)\\
\hline
$K_c=0.1694$  & 12 & 2.615(61) & 5.8(1.4)  & 2.84(44)  & 0.802(19)  &
$-$0.208(11)\\
       & 16 & 2.515(60) &	6.2(1.9)	 & 2.89(61) & 0.765(23)	 & $-$0.222(20)\\
\hline
\end{tabular*}
\vspace*{-2mm}
\end{table*}

We summarize our results for nucleon scalar matrix elements in
Table~\ref{table:two}.  For figures of three-point functions and the
of matrix elements on the quark mass, we refer to
Ref.~\cite{sigmaterm} where results on a $12^3\times 20$ lattice has been
reported.  New runs  on a $16^3\times 20$ lattice have yielded consistent
results within errors, showing that finite lattice size effects are
small.

For the $\sigma$ term matrix element our results clearly show that the
disconnected contribution is about twice larger than the connected one,
confirming, with reduced errors and also at smaller quark masses, the trend
indicated by the previous results in quenched QCD at $\beta=6.0$ obtained with
the $Z_2$ noise source\cite{z2noise} and in full QCD at $\beta$=5.4--5.6
extracted via $dm_N/dm_q$\cite{fullw}. This point is illustrated in
Fig.~\ref{fig:sigmasummary}. Averaging the results for $L=12$ and 16 we find
$\sigma_{disc}/\sigma_{conn}=2.35(46)$ at $K_c$.

theXWe estimate the physical value of $\sigma$ from $\hat ma=$ $(m_u+m_d)a/2=$
$0.0034(1)$
$(m_qa=(1/K-1/K_c)/2)$ using $m_\pi/m_\rho=0.18$ and $a^{-1}=1.46(2)$
GeV determined from $m_\rho$, yielding $\sigma=44(6)$MeV from averages
of $L=12$ and $L=16$ results.  Alternatively, extrapolating the ratio
$m_N\sigma/m_\pi^2$ linearly in $m_q$ to $K_c$ and using experimental $\pi$
and nucleon masses, we  obtain $\sigma=60(9)$MeV.
The difference between the two lattice estimations is
ascribed to a too large value of the ratio $m_N/m_\rho=1.47(3)$ at
Allowing for this uncertainty the lattice estimate shows
an encouraging agreement with a result from  dispersion
relations $\sigma=64(8)$MeV\cite{koch} or a phenomenological estimate
$\sigma\approx 45$MeV\cite{chpt} which is obtained using an extrapolation of
the dispersion result at the
Cheng-Dashen point $t=2m_\pi^2$ to $t=0$.

Systematic uncertainties in our results are scaling violation
and sea quark effects.  A potential problem is that the quark
mass for Wilson quark action decreases toward weaker couplings in
quenched QCD and that the full QCD values are a factor 2--3 smaller than those
for quenched QCD\cite{ukawa}. To what extent the $\sigma$ term is
affected by these effects should be
checked in future simulations.

The matrix element $<N|\bar{s}s|N>$ in Table~\ref{table:two} represents an
interpolated value at the physical strange quark mass  $m_{s}=0.0829(19)$
$(K_{s}=0.1648)$ determined from $m_{K}/m_{\rho}=0.64$.  We find a fairly
large value for the $y$ parameter
$y=$ $2<N|\bar{s}s|N>/<N|\bar{u}u+\bar{d}d|N>=0.66(13)$ as compared to a
phenomenological estimate $y\approx 0.2$\cite{chpt}.

The couplings $F_S$ and $D_S$ are evaluated from connected contributions. The
ratio $D_S/F_S=-0.275(16)$ is reasonably consistent with the experimental
value $D_S/F_S\approx -0.32$ deduced from the baryon octet mass
splitting.  The magnitude, however, is roughly a factor two smaller: we find
$m_\Xi-m_N=(m_s-\hat m)2F_S=0.182(6)$GeV and
$-3/2 (m_\Sigma-m_\Lambda)$=$(m_s-\hat m)2D_S$=$-0.0498(30)$GeV as compared to
the experimental values 0.379GeV and $-0.116$GeV.  We suspect that the bulk of
TheXat $\beta=5.7$.

\begin{figure}[t]
\psfig{file=delq.epsf,height=6cm,width=7.5cm,angle=-90}
\vskip -1.0cm
\caption{Quark contribution to the proton spin as a function of up and down
quark mass.}
\label{fig:axialNM}
\vspace*{-9mm}
\end{figure}

\mysection{Axial vector matrix elements}

Statistical fluctuations in the disconnected contribution to the axial vector
matrix elements turned out to be significantly larger than those for the
scalar matrix elements, to the extent that a reliable signal was not obtained
on a $12^3\times 20$ lattice.  Our results on a $16^3\times 20$ lattice are
presented in Fig.~\ref{fig:axialNM} as a function of up and down quark mass,
where $\Delta q$ represents the matrix element of $\bar q\gamma_i\gamma_5 q$
for a proton polarized in the $i$-th direction averaged over $i=1,2,3$.  The
values for $\Delta s$ are a result of an interpolation to the physical strange
quark mass with up and down quark masses fixed.

We observe that the magnitude of the disconnected contributions, though
involving a substantial error of 50\%, is small compared to those of connected
ones. Also, the disconnected contributions increase toward small quark masses
while the connected ones show a slight decrease. Adding the two contributions
and making a linear extrapolation to $m_q=0$, we find
$\Delta{u}=0.638(54)$ and
$\Delta{d}=-0.347(46)$, in which the
disconnected contributions are
$-0.119(44)$, and $\Delta{s}=-0.109(30)$ at the physical strange quark mass.
For the fraction of proton spin carried by quarks, we then obtain
$\Delta\Sigma =\Delta{u}+\Delta{d}+\Delta{s}=0.18(10)$.   These values are
amazingly similar to the results $\Delta{u}=0.80(4)$, $\Delta{d}=-0.46(4)$,
$\Delta{s}=-0.13(4)$ and $\Delta{\Sigma}=0.22(10)$ reported in
Ref.~\cite{elliskarliner} from a reanalysis of three experimental
data\cite{protonspin}.

contributions are  $F_{A}=0.382(18)$  and
$D_{A}=0.607(14)$ in the chiral limit. The ratio
$F_{A}/D_{A}=0.629(33)$ shows a good agreement with the experimental value
$F_{A}/D_{A}=0.58(5)$ obtained from data for hyperon $\beta$ decay\cite{fd}.
The magnitude, however, are about $25\%$
smaller than the experimental values  $F_{A}=0.47(4)$ and
$D_{A}=0.81(3)$\cite{fd},
which we ascribe to scaling violation.

\vspace*{2mm}
This work is supported in part by the Grant-in-Aid of the Ministry of
(No.~06NP0601, No.~06640372, No.~05-7511).

\end{document}